# LIFETIME TESTING 700 MHZ RF WINDOWS FOR THE ACCELERATOR PRODUCTION OF TRITIUM PROGRAM

K. A. Cummings, M. D. Borrego, J. DeBaca, J. S. Harrison, M. B. Rodriguez, D. M. Roybal,
W. T. Roybal, S. C. Ruggles, P. A. Torrez, LANL, Los Alamos, NM 87545, USA
G. D. White, Marconi Applied Technologies, Chelmsford, England

*Abstract*

Radio frequency (RF) windows are historically a point where failure occurs in input-power couplers for accelerators. To understand more about the reliability of high power RF windows, lifetime testing was done on 700 MHz coaxial RF windows for the Low Energy Demonstration Accelerator (LEDA) project of the Accelerator Production of Tritium (APT) program. The RF windows, made by Marconi Applied Technologies (formerly EEV), were tested at 800 kW for an extended period of time. Changes in the reflected power, vacuum, air outlet temperature, and surface temperature were monitored over time. The results of the life testing are summarized.

## 1 INTRODUCTION

A lifetime window test stand has been set up to learn more about the failure mechanisms and lifetime of RF windows for the Accelerator Production of Tritium (APT) program. Two 700 MHz RF windows prototypes were ordered from the vendor, Marconi Applied Technologies, and were conditioned in less than 20 hours and then tested to 800 kW for 4 hours. Based on these test results, 14 additional windows were ordered for window life time testing. Due to problems encountered with the initial high power testing of these additional windows, not as much life time test data was collected as planned. Two windows were tested for 245 hours and 155,513 kW hours on the test stand. The other windows that were life tested for less time encountered various problems. This paper summarizes the life data obtained on the two windows which ran for an extended period of time and the problems encountered with the other windows.

## 2 EXPERIMENTAL APPARATUS AND PROCEDURE

### 2.1 Window Geometry

The windows are a coaxial geometry using an AL995 alumina ceramic. As shown in Figure 1, the window assembly consists of half height WR 1150 waveguide on the vacuum side, a T-Bar transition into the coaxial region of the alumina ceramic, and another T-Bar to transition to full height WR 1500 waveguide on the air side. The configuration shown in Figure 1 is referred to as a right handed configuration. The air side waveguide may be rotated 180 degrees around the axis of the coax to yield a left handed configuration. Both the right handed and the left handed configurations are used to feed power to the Coupled Cavity Drift Tube Linac (CCDTL) on the APT accelerator. The vacuum side waveguide is copper plated stainless steel. The inner and outer coax is copper and the air side waveguide is aluminum.

The windows are cooled with 40 CFM of air cooling and 2 GPM of water cooling. The air inlet is at the air side T bar. The air flows through the T bar on the air side, down the inner conductor, and then exits the inner conductor through a series of exit holes and flows across the alumina ceramic. The air exits the coaxial region through a series of exit holes on the outer conductor and is vented to the ambient environment. The water circuit cools the vacuum side T bar region and the vacuum waveguide.

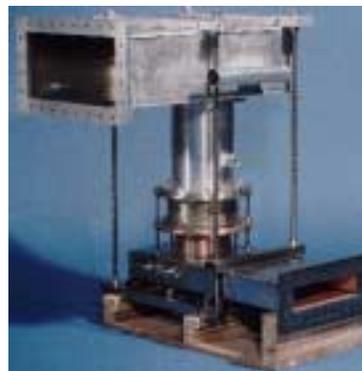

Figure 1: EEV 700 MHz RF Window.

### 2.2 Test Stand Description

The test stand was designed to test up to eight windows in series into a matched load. The windows are set up in pairs in a back to back configuration with a section of half height WR 1150 waveguide between them. Each window pair is isolated by a pair of waveguide switches. In the event of a window problem the affected pair can be isolated and the RF testing can continue. The test stand is capable of passing 1 MW CW RF power. The test stand is configured so it may be operated remotely because it is located in the beam tunnel for the Low Energy Demonstration Accelerator (LEDA) and access is not permitted when LEDA is producing beam.

### 2.3 Diagnostic Equipment

The RF window test stand includes many diagnostics. The vacuum pressure is measured at three places in each window pair and provides an interlock to the RF power.

The pressure at which the interlock is set may be adjusted during testing. Each RF window has two fiber optic arc detectors; one on the air side and one on the vacuum side. The arc detectors also are interlocked to the RF power and will shut off the RF power for 1.6 seconds upon detection of an arc. An analog voltage is displayed corresponding to the intensity of light detected. This voltage is used to detect a window that may not be arcing, but instead displays a low intensity glow. The water and air inlet and exit temperatures are monitored and are interlocked with the RF power. In addition, the surface temperature of the outer coax in the region of the ceramic is also monitored and recorded. The positions of the waveguide switches are also interlocked. If the position of a waveguide switch is changed during testing, the RF power is shut off. In addition if the switches in a given pair are not set in the same position, the RF power is disabled. A LabView® program is used to record and archive the outer coaxial and air outlet temperatures, vacuum pressure, number of arcs, and RF power. A DataHighway® system is used to link information from the window test stand in the beam tunnel to information at the klystron used to provide RF power to the test stand. An EPICS interface has been set up to remotely operate the test stand and the klystron and to archive the data.

## 3 EXPERIMENTAL RESULTS

### 3.1 Initial High Power Acceptance Test Results

When the windows first arrive at LANL they undergo high power acceptance testing. This testing consists of conditioning the windows to 800 kW and then running at 800 kW for 4 hours. Once windows pass the high power acceptance tests, they are moved to the life time test stand. This subsection summarizes the results of the initial high power acceptance testing.

Fourteen additional windows were ordered from Marconi. Thus far, a total of nineteen windows have been received at LANL. Six windows have passed the high power acceptance tests. Of the six windows, two developed vacuum leaks and two were sent back to Marconi to have additional water cooling added in the area of the vacuum side T bar because of high temperatures in this region. The remaining two windows, serial #12 and #13, are on the life time test stand at this time. Of those that did not pass initial high power testing, three windows arced excessively and were sent back to Marcnoi. Four windows showed glowing and/or heating problems. All four of these windows were grit blasted after the initial tests. Two of the four then passed the high power acceptance tests, however, they later developed vacuum leaks. The third of the four windows broke after grit blasting and the fourth window was sent back to Marcnoi for additional cooling modifications. Another window developed a vacuum leak on a Conflat® flange. This information is summarized in Table 1.

Table 1: Summary of High Power Acceptance Testing.

| Description | Quantity of Windows |
|---|---|
| Passed the High Power Tests | 2 |
| Arced | 3 |
| Glowed and/or Heated | 4 |
| Broken due to Air Cooling problem | 2 |
| Developed Vacuum leaks | 3 |
| Damaged in shipping | 2 |

### 3.2 Lifetime Window Test Results

Currently the RF windows have undergone 408.14 MW hours of life time testing. The breakdown for the total number of MW hours is summarized in Table 2. Serial #04 and #06 were removed in April and sent back to the vendor to have the additional cooling modifications added in the T-bar region. Serial #09 and #11 both developed vacuum leaks in the T-bar region and were sent back to the vendor as well. Thus, the majority of the life time test data has been acquired on Serial #12 and #13.

Table 2: Cumulative MW-Hours from Life Tests

| MW-Hours | Window Serial Number | | | | | |
|---|---|---|---|---|---|---|
| | 04 | 06 | 09 | 11 | 12 | 13 |
| Month & Year 11-99 | 16.77 | 16.77 | 16.77 | 16.77 | | |
| 12-99 | 0 | 0 | 0 | 0 | | |
| 1-00 | 0.38 | 0.38 | 0.38 | 0.38 | | |
| 2-00 | 0 | 0 | 0 | 0 | | |
| 3-00 | 14.26 | 14.26 | | | 38.36 | 38.36 |
| 4-00 | | | | | 43.73 | 43.73 |
| 6-00 | | | | | 41.21 | 41.21 |
| 7-00 | | | | | 18.19 | 18.19 |
| 8-00 | | | | | 14.03 | 14.03 |
| Total | 31.41 | 31.41 | 17.15 | 17.15 | 155.51 | 155.51 |

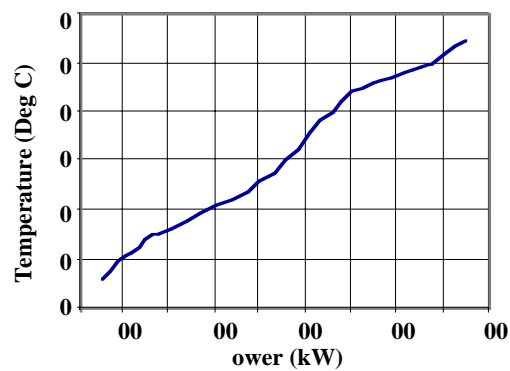

Figure 2: Steady State Temperature vs. Power.

The steady state temperature of the outer coaxial surface at various power levels are shown in Figure 2. The temperatures shown are from windows that did not show any arcing, glowing or heating problems. During the high power acceptance testing the outer coaxial surface temperatures of the windows are compared to this data as a tool to indicate any problems.

The variations in the coaxial surface temperature, air outlet temperature, amount of arcing, and vacuum pressure are examined over time. As shown in Figure 3, the coaxial surface temperature is plotted at various power levels during each month for serial numbers 12 and 13. As expected from Figure 2, the average temperature increases with the average power, however, there is no correlation of temperature, either increasing or decreasing, with time. Figure 4 shows no correlation between the air outlet temperature and test time; however, the noticeable increase in the air out temperature with power can be easily seen. The vacuum pressure remained very constant throughout the testing and there were no significant number of arcs.

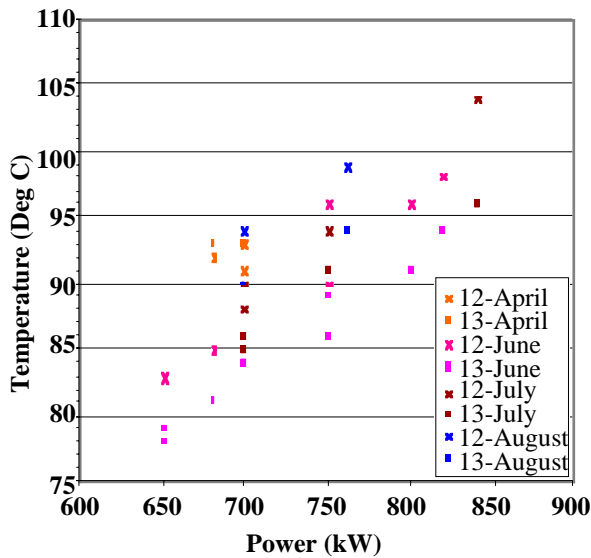

Figure 3: Average Coaxial Surface Temperatures.

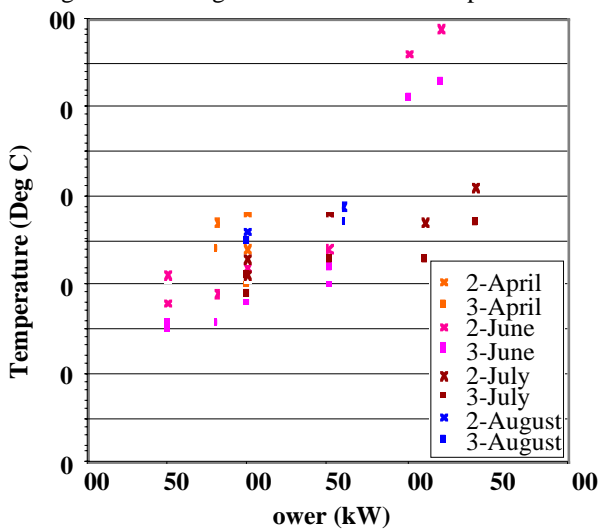

Figure 4: Average Air Out Temperatures.

## 4 DISCUSSION

The problems encountered in the initial high power testing include an air cooling problem, shipping damage, vacuum leaks, arcing, and glowing and/or heating. The air cooling problem was caused by inadequate air flow due to a mis-adjusted orifice. This cooling problem is easily explainable and fixed. The shipping damage is also easily explained. The interesting problems are the vacuum leaks, glowing and/or heating, and arcing. The vacuum leaks in the region of the T-bar are thought to be unrelated to the glowing or the grit blasting. The leaks are thought to be due to high stresses caused by a large thermal gradient in the region of the T-bar. This problem was corrected by adding a copper insert and additional cooling in the region of the T bar. One window had a vacuum leak on a Conflat® flange which was due to a poor weld. The glowing and/or heating and arcing are caused by surface contaminates and/or defects[1]. The glowing and heating are grouped together because if one problem is seen, the other one usually accompanies it. The glow is a blue or purple color and the window temperatures are 10 to 20 degrees Celsius higher. The temperature rise and the optical emission are indicators multipacting[2,3]. If a significant glow occurs, an analog voltage is indicated on the arc detectors. The intensity of the glow increases with RF power level. Windows that are arcing usually exhibit a very good vacuum and do not display higher temperatures than normal.

Given the current amount of life testing, no definitive conclusions can be made about the lifetime of the window because there were no trends in the surface and air temperatures, the amount of arcing, or vacuum pressure with time. Some improvements were made in the thermal mechanical design of the windows by adding additional cooling. Constructing the vacuum side waveguide out of copper would be a design improvement because the increase in conductivity would improve the thermal mechanical properties of the window.

## 5 CONCLUSIONS

Variations in the manufacturing and surface preparation processes have led to a low testing success rate for the initial high power testing. Other problems encountered (i.e., air cooling problems, shipping damage, vacuum leaks) can be corrected and/or fixed. Additional life test data is needed to determine the life time of the windows

## REFERENCES

[1] K. A. Cummings, "Theoretical Predictions and Experimental Assessments of the Performance of Alumina RF Windows", Ph.D. Thesis, University of California, Davis, CA, June 1998.
[2] Y. Saito, N. Matuda, S. Anami, "Breakdown of Alumina RF Windows", Rev. Sci. Instrum., Vol. 60, No. 7, July 1989.
[3] Y. Saito, N. Matuda, S. Anami, A. Kinbara, G. Horikoshi, J. Tanaka, "Breakdown of Alumina RF Windows", IEEE Trans. on Elect. Insul. , Vol. 24, No. 6, Dec 1989.